\documentclass[aps,prl,twocolumn,preprintnumbers,superscriptaddress,nofootinbib]{revtex4}
\usepackage{graphicx}
\usepackage{epstopdf}
\usepackage{amsmath}
\usepackage{amsfonts}
\usepackage{amssymb}
\usepackage{appendix}
\usepackage{enumerate}
\usepackage{comment}
\usepackage{bbold}
\usepackage[shortlabels]{enumitem}
\usepackage{color}
\usepackage{slashed}
\usepackage{subfigure}
\usepackage{setspace}
\usepackage{footnote}
\usepackage{lipsum}
\usepackage{multirow}
\usepackage[colorlinks = true,
            linkcolor = blue,
            urlcolor  = blue,
            citecolor = blue,
            anchorcolor = blue]{hyperref}
\usepackage[capitalize]{cleveref}
\usepackage{braket}
\usepackage[compat=1.1.0]{tikz-feynman}
\usepackage{multirow}
\usepackage{physics}
\usepackage{feynmp-auto}
\usepackage[normalem]{ulem}

\def\bar{\overline}

\begin{document}

\singlespacing

\preprint{FERMILAB-PUB-22-648-T}
\preprint{NUHEP-TH/22-08}

\title{Addressing the Short-Baseline Neutrino Anomalies \\ with Energy-Dependent Mixing Parameters}

\author{K.~S.~Babu}
\email{kaladi.babu@okstate.edu}
\affiliation{Department of Physics, Oklahoma State University, Stillwater, OK, 74078, USA}
\author{Vedran Brdar} 
\email{vbrdar@fnal.gov}
\affiliation{Northwestern University, Department of Physics \& Astronomy, 2145 Sheridan Road, Evanston, IL 60208, USA}
\affiliation{Theoretical Physics Department, Fermilab, P.O. Box 500, Batavia, IL 60510, USA}
\author{Andr\'{e} de Gouv\^{e}a} 
\email{degouvea@northwestern.edu}
\affiliation{Northwestern University, Department of Physics \& Astronomy, 2145 Sheridan Road, Evanston, IL 60208, USA}
\author{Pedro A.~N.~Machado}
\email{pmachado@fnal.gov}
\affiliation{Theoretical Physics Department, Fermilab, P.O. Box 500, Batavia, IL 60510, USA}

\begin{abstract}
Several neutrino experiments have reported results that are potentially inconsistent with our current understanding of the lepton sector. A candidate solution to these so-called short-baseline anomalies is postulating the existence of new, eV-scale, mostly sterile neutrinos that mix with the active neutrinos. This hypothesis, however, is strongly disfavored once one considers all neutrino data, especially those that constrain the disappearance of muon and electron neutrinos at short-baselines. Here, we show that if the sterile--active mixing parameters depend on the energy-scales that characterize neutrino production and detection, the sterile-neutrino hypothesis may provide a reasonable fit to all neutrino data. The reason for the improved fit is that the stringent disappearance constraints on the different elements of the extended neutrino mixing matrix are associated to production and detection energy scales that are different from those that characterize the anomalous LSND and MiniBooNE appearance data. We show, via a concrete example, that secret interactions among the sterile neutrinos can lead to the results of interest. 
\end{abstract}

\maketitle

\textbf{Introduction.}
In the era of the ground-breaking experiments that have led to the discovery of neutrino oscillations and nonzero neutrino masses \cite{superk,sno,kamland}, there have been several experiments whose findings continue to challenge the standard three-massive-active-neutrinos paradigm. We refer to these as the short-baseline anomalies. The first hint for neutrino flavor-change at baselines that are inconsistent with what is currently known about neutrino masses -- in hindsight, the baseline turned out to be  too short -- came from the LSND experiment \cite{LSND,LSND:2001aii}. Data from LSND can be interpreted as evidence for $\bar{\nu}_{\mu}\to \bar{\nu}_e$ oscillations. The MiniBooNE experiment \cite{MiniBooNE:2008hfu}, originally designed to test the oscillation-interpretation of the LSND data, is consistent with an excess of $\nu_e$-like events at low energies \cite{MiniBooNE:2008yuf,MiniBooNE:2018esg,MiniBooNE:2020pnu}. MiniBooNE data can be interpreted as evidence for $\nu_{\mu}\to \nu_e$ and $\bar{\nu}_{\mu}\to \bar{\nu}_e$ oscillations. In parallel, short-baseline reactor neutrino experiments reported a deficit of electron antineutrinos from reactor sources \cite{Mention:2011rk,Huber:2011wv}. 
These data are consistent with $\bar{\nu}_{e}$ disappearance but recent data and analyses indicate the so-called reactor anomaly may turn out to be the result of a mis-estimation of the flux of antineutrinos from nuclear reactors \cite{Giunti:2021kab,Berryman:2021yan}. Finally, there is the so-called gallium anomaly \cite{SAGE:1998fvr,Abdurashitov:2005tb,Kaether:2010ag}, which is consistent with short-baseline $\nu_{e}$ disappearance and for which the BEST experiment \cite{Barinov:2021asz} has  recently provided supporting evidence. In this paper we will focus on the LSND and MiniBooNE anomalies; the latter is still in great focus as it is currently being tested by the Short-Baseline Neutrino (SBN) Program at Fermilab \cite{Machado:2019oxb,MicroBooNE:2021ktl,MicroBooNE:2021zai}. 

The introduction of eV-scale mostly sterile neutrinos can successfully explain the LSND and MiniBooNE results \cite{MiniBooNE:2020pnu,Brdar:2021ysi,MiniBooNE:2022emn}. However, this hypothesis is not favored once all available neutrino data are considered in tandem \cite{Dentler:2018sju,Gariazzo:2017fdh,Diaz:2019fwt}. In particular, in the regions of parameter space where one can fit the data from LSND and MiniBooNE and satisfy constraints from solar and reactor neutrino data, there is strong tension with muon neutrino disappearance searches performed by, especially, the IceCube \cite{IceCube:2020phf} and MINOS \cite{MINOS:2017cae} experiments. Here, we argue that this tension can be significantly alleviated if one allows the active--sterile mixing parameters to depend on the energy scales associated to neutrino production and detection. We further demonstrate that secret interactions between the sterile neutrinos and a relatively light new $U(1)$ gauge boson can lead to the effect of interest.

\textbf{Energy-dependent mixing parameters.}
New neutrino interactions lead to energy-dependent neutrino mixing parameters. In Ref.~\cite{Babu:2021cxe}, we demonstrated that, if the new particles are light enough, this energy dependency can lead to observable effects in oscillation experiments. To introduce the effect, one can focus on the probability that a neutrino produced with energy $E_{\nu}$ as a $\nu_{\alpha}$ will be detected as a $\nu_{\beta}$, $\alpha,\beta=e,\mu,\tau$, after propagating a distance $L$, given by
\begin{align}
P_{\alpha\beta}=\sum_{j,k} U_{\alpha j}^* U_{\beta j} U_{\alpha k} U_{\beta k}^* \, e^{-i \frac{m_j^2-m_k^2}{2 E_\nu}L}\,,
\label{eq:P_osc}
\end{align} 
where $j,k=1,2,3,\ldots$, and $m_j$ are the neutrino masses, associated to the mass-eigenstates $\nu_j$. In Eq.~(\ref{eq:P_osc}), the elements of the mixing matrix $U_{\alpha j}$ ($\alpha=e,\mu,\tau$) are proportional to the couplings between the $W$-boson, the charge-lepton $\ell_{\alpha}$, and the neutrino mass-eigenstate $\nu_j$. Quantum mechanical effects lead the magnitude of the entries of $U$ to ``run,'' i.e., $U$ depends on the momentum-scale associated to the process that leads to neutrino production or detection. Including these renormalization-group-running (RG) effects, the products  $U_{\alpha j}^* U_{\alpha k}$ in Eq.~(\ref{eq:P_osc}) should be evaluated at the production energy scale while the products $U_{\beta j} U_{\beta k}^*$ in Eq.~(\ref{eq:P_osc}) should be evaluated at the relevant detection energy scale. 

The key point is that processes of neutrino production and detection do not necessarily correspond to the same energy scale, even at a single experiment. In experiments like LSND and MiniBooNE, neutrinos are chiefly produced in pion decay and therefore the relevant production energy scale is the pion mass. In contrast, neutrino detection typically occurs via neutrino charged-current scattering on nucleons, so the relevant detection energy scale is a function of $E_\nu$ and the nucleon mass, $m_N$. In this work (as in Ref.~\cite{Babu:2021cxe}), for the detection energy scale, we choose $\sqrt{2 E_\nu^2 m_N/(2 E_\nu+m_N)}$, the square root of the mean value of the Mandelstam variable $t$. The final ingredients in Eq.~(\ref{eq:P_osc}) are the neutrino mass-squared differences $m_j^2-m_k^2$. While the mass parameters in general also depend on the energy scale, in neutrino oscillations they correspond to the physical, on-shell masses, as we argued in \cite{Babu:2021cxe} (see also \cite{Volobuev:2017izt}). On-shell masses, of course, do not run.

In \cite{Babu:2021cxe}, we presented different new-physics scenarios, some of which were associated to the  the origin of nonzero neutrino mass, where RG effects are significant for the energy scales associated to neutrino experiments. We also discussed, for the three-massive-active-neutrinos paradigm, phenomenological implications for the long-baseline T2K \cite{T2K:2019ird} and NOvA \cite{NOvA:2019cyt} experiments as well as the flavor composition of astrophysical neutrinos observed in IceCube \cite{IceCube:2015rro} (see also \cite{Bustamante:2010bf}). Here, we extend these ideas to the active--sterile neutrino mixing sector and illustrate how the short-baseline anomalies can be resolved by this mechanism.

\textbf{The model.}
We add to the Standard Model (SM) particle content two SM gauge-singlet fermions $N$ and $N'$ (sterile neutrinos) and a SM gauge-singlet scalar $S$, and assume
\begin{align}
\mathcal{L}\supset \frac{C_\alpha}{\Lambda} \bar{L}_{\alpha} H S N + M N' N + \text{h.c.}\,,
\label{eq:Lag}
\end{align}
where $L_{\alpha}$, $\alpha=e,\mu,\tau$, and $H$ are, respectively, the SM lepton and Higgs doublets, $C_{\alpha}$ are dimensionless coupling constants, $\Lambda$ is the energy scale that characterizes the dimension-five effective operator in Eq.~(\ref{eq:Lag})\footnote{We will not discuss the physics that leads to such a term at low energies. It is enough to assume it involves fields that are at or above the electroweak symmetry breaking scale so that this  physics does not contribute to the running of mixing parameters at the energy scales of interest.} and $M$ is the (Dirac) sterile neutrino mass. We assume that  $S$ acquires a nontrivial vacuum expectation value. In the context of the short baseline anomalies, $\mu_{\alpha} \equiv C_{\alpha}\langle H \rangle \langle S \rangle /\Lambda \sim 0.01-0.1$~eV and $M\sim 1$~eV are required. After spontaneous symmetry breaking, Eq.~(\ref{eq:Lag}) adds to the neutrino mass matrix, expressed here in the extended weak-eigenstate basis ($\nu_{\alpha},N,N'$):
\begin{align}
M_\nu=
\begin{pmatrix}
\cross & \cross & \cross & \mu_e & 0 \\
\cross & \cross & \cross & \mu_\mu & 0 \\
\cross & \cross & \cross & \mu_\tau & 0 \\
\mu_e & \mu_\mu & \mu_\tau & 0 & M \\
0 & 0 & 0 & M & 0 
\end{pmatrix}\,.
\label{eq:massmatrix}
\end{align}
The upper-left $3\times 3$ submatrix in \cref{eq:massmatrix} (whose elements are denoted with `$\cross$') is given by $U^* \,\text{diag}(m_1,m_2,m_3) \, U^\dagger$, where $m_i$ are the mostly-active neutrino masses.\footnote{We are agnostic regarding the physics that leads to the mostly-active neutrino masses. It is unrelated to the presence of $N$ and $N'$, by design. One viable scenario for active neutrino mass generation is the canonical Type-I seesaw mechanism \cite{Minkowski,Yanagida:1979as,GellMann:1980vs,Goran,Schechter:1980gr}.}
In order to address the LSND and MiniBooNE anomalies, we need mixing between the mostly sterile states and the electron and muon neutrinos, hence $\mu_e$ and $\mu_\mu$ need to be nonzero. 
Given that data are mostly silent regarding mixing with $\nu_{\tau}$, for simplicity, we set $\mu_\tau=0$. 
For concrete numerical computations, we assume the so-called normal mass ordering and that $m_1$ vanishes. The nonzero masses are derived from the mass-squared differences, whose values agree with the best-fit values in  \cite{Esteban:2020cvm}. 
For the elements of the $3\times 3$ ``active'' neutrino mixing matrix $U$, we took the best-fit values for the mixing angles from \cite{Esteban:2020cvm} and set the CP-violating phases to zero. 

 The model is designed to minimize the running of the mostly-active mixing parameters. 
 In particular, \cref{eq:massmatrix} is such that, at leading order in $\mu_{\alpha}/M$, the mostly-active neutrino masses and mixing parameters do not depend on $\mu_{\alpha}$ or $M$. This is easy to understand. In the absence of mostly-active neutrino masses (`$\cross$' in \cref{eq:massmatrix}), Eq.~(\ref{eq:Lag}) is invariant under a $U(1)$ global lepton-number symmetry where $N$ has lepton number $+1$ and $N'$ has lepton number $-1$. In this case, one finds only one massive neutrino. 
After symmetry breaking, the left-chiral projection of the massive state -- a Dirac fermion -- is a linear combination of the active neutrino states and $N$.

Upon diagonalizing $M_{\nu}$, the five mass eigenvalues are, to leading order in $\mu_\alpha/M$, $m_1,m_2,m_3,M,-M$ (the negative sign is not physical). 
Since $M\gg \mu_{\alpha},m_i$, the heaviest two states are mostly-sterile neutrinos with approximately identical masses and mixing with the active neutrinos. 
As far as short-baseline oscillations are concerned, these can be treated as one effective state $\nu_4$.  
The effective parameters characterizing $\nu_4$ mixing with electron -- $\theta_{14}$ -- and muon -- $\theta_{24}$ -- neutrinos read \cite{Deppisch:2015qwa,Berryman:2015nua,Dentler:2018sju} 
\begin{align}
\text{tan} \,\theta_{14}&\simeq \frac{\mu_e}{M},&  \text{tan} \,\theta_{24}&\simeq \frac{\mu_\mu}{M}\,. 
\label{eq:angles}
\end{align}

While here the $\mu_\alpha$ parameters do not run below the electroweak scale, the mixing angles in \cref{eq:angles} are energy-dependent as long as $M$ is the subject of relevant quantum corrections. In order to induce a significant change of $M$ across the energy scales of interest, we introduce a new $U(1)'$ gauge interaction between the sterile neutrinos and a new gauge boson $Z'$,
\begin{align}
\mathcal{L}\supset g' \bar{N} \slashed{Z'} N - g' \bar{N}' \slashed{Z'} N'\,, 
\label{eq:gauge_int}
\end{align}
where we assign equal and opposite $U(1)'$ charges to $N$ and $N'$ while none of the SM particles are charged. For previous studies involving sterile neutrino interactions with novel gauge bosons see, for example, \cite{deGouvea:2015pea,Cherry:2016jol,Bertuzzo:2018ftf, Chu:2018gxk,Berbig:2020wve}. Since there are two sterile neutrinos with opposite $U(1)'$ charges, the theory is anomaly free. The new scalar field $S$ is also charged under $U(1)'$ and we choose its charge to be such that $NS$ is a $U(1)'$ singlet and the dimension-five term in Eq.~(\ref{eq:Lag}) is gauge invariant. When $S$ acquires an expectation value, the $Z'$ vector boson acquires a mass $M_{Z'}$ that we assume to be of order a few MeV. 

The new gauge interaction from \cref{eq:gauge_int} introduces a scale-dependence to the parameter $M$,
\begin{align}
M(\mu)=M(\mu_0) \left(1 - \frac{5g'(\mu_0)^2}{24 \pi^{2}}\ln(\frac{\mu}{\mu_0})\right)^{9/4}\,,
\label{eq:RG}
\end{align}     
where $\mu$ is the energy-scale where $M$ is being evaluated, $\mu_0$ is a reference value, which we associate with low energies, and $g'(\mu)$ is the coupling constant at the energy-scale $\mu$. We included contributions from $N(1),\,N'(-1)$, and $S(-1)$ for the running of $g'$, where the $U(1)'$ charges of the fields are indicated in parenthesis. As $\mu$ increases, $M(\mu)$ decreases and the mixing angles in \cref{eq:angles} increase with energy. This is precisely the effect we require in order to improve the consistency of the sterile-neutrino solution to the LSND and MiniBooNE anomalies. Note that the decrease of $M(\mu)$ with energy has its analogue in QED: the electron mass gets smaller as the energy scale increases (see, for example, \cite{Schwartz:2014sze}).

As one runs towards lower energies, the running, roughly speaking, ``stops'' once the virtual particles are heavy relative to the energy scale. The massive particle here is the $Z'$. Throughout, we fix $M_{Z'}=5$~MeV. We checked that an MeV-scale $Z'$ that couples only to
sterile neutrinos does not run into any experimental or observational bounds, even after one allows for active--sterile neutrino mixing. In particular, constraints are significantly weaker relative to the case where the new $Z'$ couples directly to active neutrinos \cite{Blinov:2019gcj}.

\textbf{Running of sterile neutrino mixing angles.}
The excess of electron (anti)neutrinos at LSND and MiniBooNE can be explained by eV-scale sterile neutrinos. These induce neutrino oscillations for $L/E_\nu={\cal O}(1~{\rm m/MeV})$, provided there is mixing with both electron and muon neutrinos. The $\nu_{\mu}\to \nu_e$ oscillation probability, assuming a short baseline ($L/E_\nu\ll 1~{\rm km/MeV}$) and one eV-scale mostly sterile neutrino $\nu_4$ with mass $m_4$, reads 
\begin{align}
P_{\mu e}= \sin^2 2\theta_{\mu e} \sin^2 \frac{\Delta m_{41}^2 L}{4 E_\nu}\,,
\label{eq:short_baseline_approx}
\end{align}
where $\sin^2 2\theta_{\mu e}= 4 |U_{\mu 4}|^2 |U_{e 4}|^2$ and $m_4^2-m_1^2\equiv \Delta m_{41}^2$. In the language of the mixing angles introduced in \cref{eq:angles}, $U_{e 4}=\sin \theta_{14}$ and $U_{\mu 4}=\sin \theta_{24} \cos \theta_{14}$. 

For the $\Delta m^2_{41}$ values of interest, $\theta_{14}$ is strongly constrained by experiments at the MeV scale, namely reactor and solar neutrino experiments \cite{Kopp:2013vaa}. The results from reactor experiments, however, are very dependent on flux estimates \cite{Giunti:2021kab} so we will chiefly focus on constraints from solar experiments \cite{Goldhagen:2021kxe}. There are, nonetheless, uncertainties from solar physics. These are captured by the existence of various solar models that translate into different bounds. Here, we will consider limits on $\theta_{14}$ from two such models -- GS98 and AGSS09 \cite{Vinyoles:2016djt}. The latter translates into the strongest bound on $\theta_{14}$.

For the $\Delta m^2_{41}$ values of interest, the strongest bounds on $\theta_{24}$ come from the MINOS \cite{MINOS:2017cae} experiment. Another powerful probe is the IceCube measurement of the disappearance of TeV-scale muon neutrinos propagating through the Earth; in the presence of eV-scale sterile neutrinos, this effect is strongly enhanced by matter effects \cite{Nunokawa:2003ep,IceCube:2020phf}. 
While $\theta_{14}$ is more strongly constrained at the MeV scale, limits on $\theta_{24}$ come from experiments where the neutrino detection process is characterized by neutrino energies at or above 1~GeV. 

Although disappearance data require $\sin^2\theta_{14},\sin^2\theta_{24}$ to be small, the MiniBooNE and LSND data require both $\sin^2\theta_{14},\sin^2\theta_{24}$ to be bounded from below (see \cref{eq:short_baseline_approx}). In a nutshell, this is the source of the tension that ultimately disfavors the eV-scale sterile neutrino hypothesis as the solution to the short-baseline anomalies. If one allows for the possibility that the sterile--active mixing parameters depend on the energy-scale, however, the tension can be alleviated. 

\begin{figure}[t]
	\centering
	\includegraphics[scale=0.5]{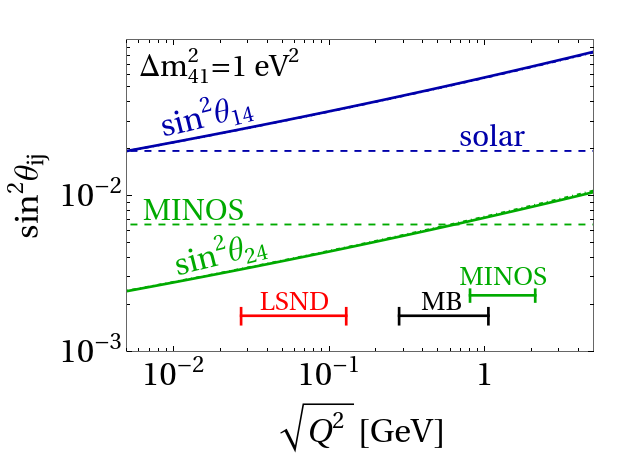} 
	\caption{Renormalization group evolution of $\sin^2\theta_{14}$ and $\sin^2\theta_{24}$. Here, $g'(\mu_0)=1.41$ and $\Delta m_{41}^2=1\,\text{eV}^2$. See text for details. The dashed horizontal lines indicate the constraints on $\sin^2\theta_{14}$ and $\sin^2\theta_{24}$ in the absence of quantum corrections. The energy scales associated to neutrino detection at LSND, MiniBooNE, and MINOS are also indicated.}
	\label{fig:3}
\end{figure}

When it comes to experiments sensitive to $\theta_{24}$, including IceCube and MINOS, 
 the production energy scale is also, like for LSND and MiniBooNE, the pion mass, while the detection energy scales are somewhat higher: the neutrino energies are $E_\nu\sim 0.1$~GeV for LSND, $E_\nu\sim 0.8$~GeV for MiniBooNE, $E_\nu\sim 3$~GeV for MINOS, and $E_\nu\sim 1,000$~GeV for IceCube. The range of detection energy scales at LSND, MiniBooNE, and MINOS are indicated in \cref{fig:3}.
 
\cref{fig:3} depicts the value of $\sin^2\theta_{14}$ in blue and $\sin^2\theta_{24}$ in green as a function of $\mu=\sqrt{Q^2}$, the relevant energy scale associated to a given production or detection process. 
We choose $\sin^2\theta_{14}(\mu=5~\rm MeV)=0.0195$, indicated by the blue, horizontal, dashed line, so it saturates the two-sigma solar bound assuming the AGSS09 solar model and $\sin^2\theta_{24}(\mu=5~\rm MeV)=0.0025$, depicted by the green, horizontal, dashed line, so that when the mixing runs to higher scales it satisfies the MINOS constraint at 90\% CL for $\Delta m^2_{41}=1$~eV$^2$. 
We choose $g'(\mu=5 ~\text{MeV})=1.41$ and $M_{Z'}=5$~MeV, typical of $^8$B solar neutrino energies. 
We perform the calculation for both the effective-four-neutrinos and the complete five-neutrinos picture (solid and dot-dashed lines, respectively) noting that the differences are negligible, as expected. 
In the five-neutrinos picture, $\sin^2\theta_{14}\equiv|U_{e4}|^2+|U_{e5}|^2$ while $\cos^2\theta_{14}\sin^2\theta_{24}\equiv |U_{\mu4}|^2+|U_{\mu5}|^2$. 
\cref{fig:3} illustrates our main point. 
RG effects can make the solar bound on $\sin^2\theta_{14}$ significantly weaker at LSND/MiniBooNE energy scales, by roughly a factor 4. 
For $\sin^2\theta_{24}$, the running renders the MINOS constraint only a little stronger, particularly compared to the MiniBooNE scale.
Combined, the running allows for larger $\nu_\mu\to\nu_e$ appearance given the disappearance constraints.

\textbf{Constraints on the mixing angles.}
 In order to  quantify how much RG effects can improve the likelihood of the eV-scale sterile-neutrinos hypothesis, we first consider constraints on $\theta_{14}$ and $\theta_{24}$ once RG effects are taken into account.
The solar neutrino constraints on $\theta_{14}$ discussed earlier in the context of the standard scenario also apply in the context of a running $\theta_{14}$, except that those constraints limit $\theta_{14}$ at low energies (below 10~MeV). There are also high energy constraints on $\theta_{14}$ from atmospheric neutrinos at Super-Kamiokande and
IceCube, discussed in \cite{Dentler:2018sju}. While these are not as strong as the solar constraints, since they constrain $\theta_{14}$ at higher energies, they are competitive. 

As already briefly discussed, $\sin^2\theta_{24}$ is constrained by short-baseline searches for $\nu_{\mu}$ disappearance. To discuss the energy dependency, for illustrative purposes, we make use of Eq.\,(III.7) in \cite{Babu:2021cxe}.  In the absence of CP violation, it reads
\begin{align}
P_{\mu\mu}=\cos(\theta_p-\theta_d)^2 - \sin 2\theta_p \sin 2\theta_d \sin^2 \frac{\Delta m_{41}^2 L}{4 E_\nu}\,,
\label{eq:Yuval}
\end{align}
where $\theta_p$ and $\theta_d$ are the values of the mixing angle -- here $\theta_{24}$ -- at production and detection, respectively. We can use this expression to ``map'' the energy-dependent mixing parameters into the standard energy-independent ones, quoted by the different experiments as a function of $\Delta m^2_{41}$ and obtained using the standard oscillation probability
\begin{align}
P_{\mu\mu}=1 - \sin^22\theta \sin^2 \frac{\Delta m_{41}^2 L}{4 E_\nu}\,.
\end{align}
One can see, for example, by comparing the amplitude of the oscillation, that $\sin 2\theta_p \sin 2\theta_d$ plays the role of $\sin^22\theta$ once RG effects are taken into account. For $\theta_{24}$-related bounds, $\theta_p$ is evaluated at $\sqrt{Q^2}=m_\pi\simeq 0.14$ GeV for atmospheric and accelerator neutrinos. On the other hand, $\theta_d$ is evaluated at $\sqrt{Q^2}=\sqrt{2 E_\nu^2 m_N/(2 E_\nu+m_N)}$. We take $E_\nu=3\,(10)$ GeV for MINOS (atmospheric neutrinos).  

In numerical analyses, we do not use \cref{eq:Yuval} but instead include all neutrino flavors and the RG dependent mixing matrix that comes from \cref{eq:massmatrix}. 
\cref{eq:Yuval} assumes, for example, that all mass-squared differences other than $\Delta m_{41}^2$ are zero, an approximation that is not valid for MINOS and atmospheric neutrinos. 
The combination $\sin 2\theta_p \sin 2\theta_d$, nonetheless, is a natural ``building block'' of the more general expression, rendering  \cref{eq:Yuval} useful. 
When computing constraints on the allowed values of the parameters, we also include constraints from experiments that operated at higher energies: CDHS $\nu_\mu$ disappearance \cite{VonRuden:1982fp} and NuTeV $\nu_e$ and $\nu_\tau$ appearance \cite{CCFRNuTeV:1998gjj,NuTeV:2002daf} at $\sqrt{Q^2}\approx 4.5$~GeV.
These come from zero-baseline flavor transitions arising from the mismatch between the mixing matrices at production and detection, keeping in mind both sterile and active mixing angles are subject to running effects. As discussed earlier, the running of active mixing angles is, by design, relatively suppressed.

Another important signature of the sterile-neutrino hypothesis is the disappearance of TeV-scale muon antineutrinos passing through the Earth, where the oscillation probability is enhanced due to matter effects \cite{Nunokawa:2003ep}. 
These are constrained by IceCube \cite{IceCube:2020phf}. 
The matter potential for muon (anti)neutrinos depends on the electron and neutron number density of the medium, as well as on neutrino oscillation parameters $\Delta m^2$ and $\theta$ \emph{evaluated at zero momentum-transfer}. The energy-dependent mixing parameter formalism for neutrino propagation in constant density matter (here we assume the Earth to have a constant density of $5 \, \text{g}/\text{cm}^3$ and that the number of protons and neutrons is the same) was discussed in detail in \cite{Babu:2021cxe} (see section III B).  

For a simplified two-flavor system (muon and sterile neutrino), the effective mixing angle in matter is 
\begin{align}
\sin^2 2\theta_{\rm eff} \approx \frac{\left[\Delta m^2 \sin 2\theta-2E_\nu V \sin(2\theta-2\theta_0)\right]^2}{(\Delta m^2 \cos2\theta_0-2E_\nu V)^2+(\Delta m^2 \sin2\theta)^2}\,,
\label{eq:angle_eff}
\end{align}
where $\theta_0$ denotes the mixing angle at zero momentum transfer, $V=\pm\sqrt{2}/2 G_F n_n$ (where $n_n$ is the neutron number density), the positive sign corresponding to muon antineutrinos, and, for simplicity, here we ignore the difference between production and detection scale mixing angles -- both are denoted as $\theta$. 
The familiar expression for the mixing angle in matter is reproduced for $\theta=\theta_0$, when the second term in the numerator of \cref{eq:angle_eff} vanishes. 
In a general scenario with energy-dependent mixing parameters we observe that, when the resonance condition ($\Delta m^2 \cos2\theta_0=2E_\nu V$) is satisfied, $\theta_{\rm eff}$ does not reach maximal mixing, unlike in the standard scenario. 
In practice, however, for the values of the parameters that are of interest here, this deviation from maximal mixing is rather small. 

 \begin{figure}[t]
	\centering
	\includegraphics[scale=0.5]{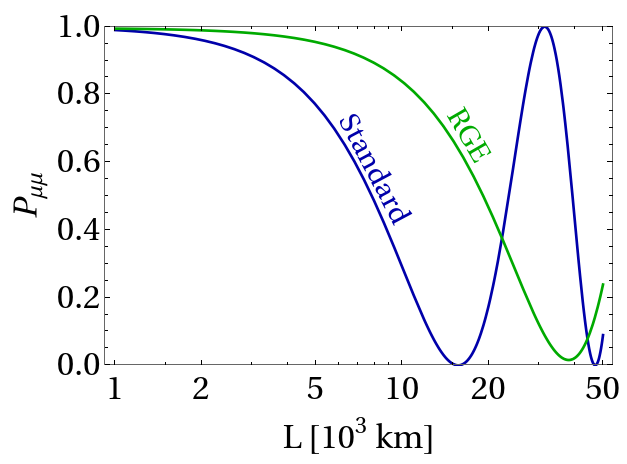} 
	\caption{
Disappearance probability of TeV-scale muon antineutrinos in the standard (blue) and energy-dependent frameworks (green), assuming only two-flavors (muon and sterile neutrino) and that the antineutrinos go  through the Earth. $\sqrt{\Delta m^2}=0.35$ eV and $E_\nu$ is set to the resonant energy. In both curves, the mixing angles are chosen such that the amplitude of $\nu_{\mu}\to\nu_e$ oscillations at MiniBooNE is $3\times 10^{-3}$. In the energy-dependent framework, relative to the standard scenario, the resonance is shifted towards larger values of $L$, often exceeding the Earth's diameter for the values of the parameters of interest. 
}
	\label{fig:2}
\end{figure}
Nevertheless, the impact of the running is significant when interpreting IceCube constraints on sterile neutrino mixing.
\cref{fig:2} depicts $P_{\mu\mu}$ as a function of the baseline $L$, for both the standard (blue) and energy-dependent (green) scenarios, assuming the neutrino energy is such that the resonance condition is met. For both curves, $P_{\mu\mu}$ reaches values that are very close to zero ($\sin^22\theta_{\rm eff}$ very close to 1). 
In \cref{fig:2}, the mixing angle and mass-squared difference are interpreted as $\theta_{24}$ and $\Delta m^2_{41}$, respectively. 
In order to establish a connection with, for concreteness, the MiniBooNE anomaly, in the standard case (blue curve), we choose 
$\sin^2\theta_{14}=0.0195$, which saturates the 2$\sigma$ C.L. bound from solar data assuming the  AGSS09 solar model \cite{Goldhagen:2021kxe}, 
and $\sin^2\theta_{24}=0.039$ such that $\sin^2 2\theta_{\mu e}=3\times 10^{-3}$  (see Eq.~(\ref{eq:short_baseline_approx})). 
Since $E_{\nu}$ is such that the resonance condition is met, the dependency on  $\Delta m_{41}^2$ in the region of interest is negligible. 
In the case of the energy-dependent scenario (green curve), we set $g'(\sqrt{Q^2}=5\,\text{MeV})=1.34$, choose the same value for $\theta_{14}$ (at low energies) as before, and fix $\sin^2\theta_{24}(\sqrt{Q^2}=5~{\rm MeV})=0.0069$ such that, at MiniBooNE, the amplitude of the oscillatory term in $P_{\mu e}$ is $3\times 10^{-3}$ (cf.~\cref{eq:Yuval2}). 
This allows an ``apples to apples'' comparison between the standard and the energy-dependent scenarios. 

The most relevant difference between the standard and new physics cases is the resonant oscillation length: $L\approx\pi/(V \sin 2\theta)$  in the standard case, versus $L\approx\pi/(V \sin 2\theta_0)$ in energy-dependent framework. 
Since $\theta_0$ is small relative to $\theta$ -- see \cref{fig:3}, keeping in mind $\theta_0$ is evaluated in the low energy region -- the resonant baseline is shifted to higher $L$ in the energy-dependent case, i.e., the first minimum of the green curve is at larger values of the propagation length. 
This relative shift weakens the standard limit from IceCube. 
For the values of $\Delta m^2_{41}$ of interest, the standard case leads to a more pronounced muon antineutrino disappearance effect inside the Earth ($2R_{\oplus}\sim 13,000$~km).

To evaluate the IceCube constraint on the energy-dependent mixing scenario, we performed the following simplified analysis.
We define the number of $\nu_\mu$ resonant events for a given model as $N_{\rm res}^{\rm model}=C \times \int d (\cos \zeta) \phi(E_\nu^{\rm res}) \sigma(E_\nu^{\rm res}) P_{\mu\mu}^{\rm model}$, where $E_{\nu}^{\rm res}$ is the neutrino energy which yields resonant oscillations in the Earth, $C$ is a constant, and the integral is over the zenith angle $\zeta$.
A point in our model would be ruled out if $N_{\rm res}^{\rm run} < N_{\rm res}^{\rm stand}$, where ``run'' and ``stand'' refers to the running and the standard $3+1$ sterile neutrino models, respectively, for the same $\Delta m^2_{41}$.
$N_{\rm res}^{\rm stand}$ is calculated using the value of $\theta_{24}$ in the $3+1$ scenario that is constrained at the $99\%$ C.L. by IceCube~\cite{IceCube:2020phf}.
We find that the IceCube bound for the running scenario is always significantly weaker than the bounds from MINOS, CDHS and NuTeV, and thus we do not consider IceCube in our combined constraints.

\textbf{Results.} 
We interpret MiniBooNE and LSND as a consequence of $\nu_{\mu}\to\nu_e$ oscillations. The 
analogue of \cref{eq:short_baseline_approx} in the energy-dependent-mixing-parameter formalism reads \cite{Babu:2021cxe}
\begin{align}
P_{\mu e}=\sin(\theta_p-\theta_d)^2 + \sin 2\theta_p \sin 2\theta_d \sin^2 \frac{\Delta m_{41}^2 L}{4 E_\nu}\,,
\label{eq:Yuval2}
\end{align}
where $\theta_p$ and $\theta_d$ are the effective mixing angle between electron and muon neutrinos, $\theta_{\mu e}$, evaluated at the production and detection scales, respectively ($\sin2\theta_{\mu e}=2|U_{e4}||U_{\mu4}|$). \cref{eq:Yuval2} contains a baseline-independent term $\sin(\theta_p-\theta_d)^2$ that leads to zero-baseline flavor transitions. For addressing the MiniBooNE and LSND data, this term is always suppressed (by at least an order of magnitude) relative to the amplitude of the oscillatory term.  This indicates that $P_{\mu e}$, for parameter values that are relevant for addressing the MiniBooNE and LSND anomalies, is well approximated by \cref{eq:short_baseline_approx} with an effective $\sin^2 2\theta_{\mu e}$.
When computing our final results, instead of making use of \cref{eq:Yuval2}, we 
perform the calculation in the full five-flavor picture and extract the effective $\sin^2 2\theta_{\mu e}$ by isolating the coefficient of the oscillatory term. 

We perform a scan of the parameter space for different models, identifying the values of $\Delta m^2_{41}$ and the (effective) $\sin^22\theta_{\mu e}$ that are consistent with the neutrino data except for those associated to the short-baseline anomalies. 
Results from $\nu_e$ appearance at MicroBooNE, estimated in \cite{Arguelles:2021meu},
are also included in our computations\footnote{MicroBooNE is also sensitive to $\nu_{\mu}$ and $\nu_e$ disappearance \cite{Arguelles:2021meu}. Such limits are relatively less relevant for our analyses and have not been included.}.
In parallel, we consider the region of the same parameter space preferred by MiniBooNE or LSND. We use the comparison of these two regions in order to gauge how well a specific model can accommodate all the data. 

\begin{figure}[t]
	\centering
	\includegraphics[scale=0.375]{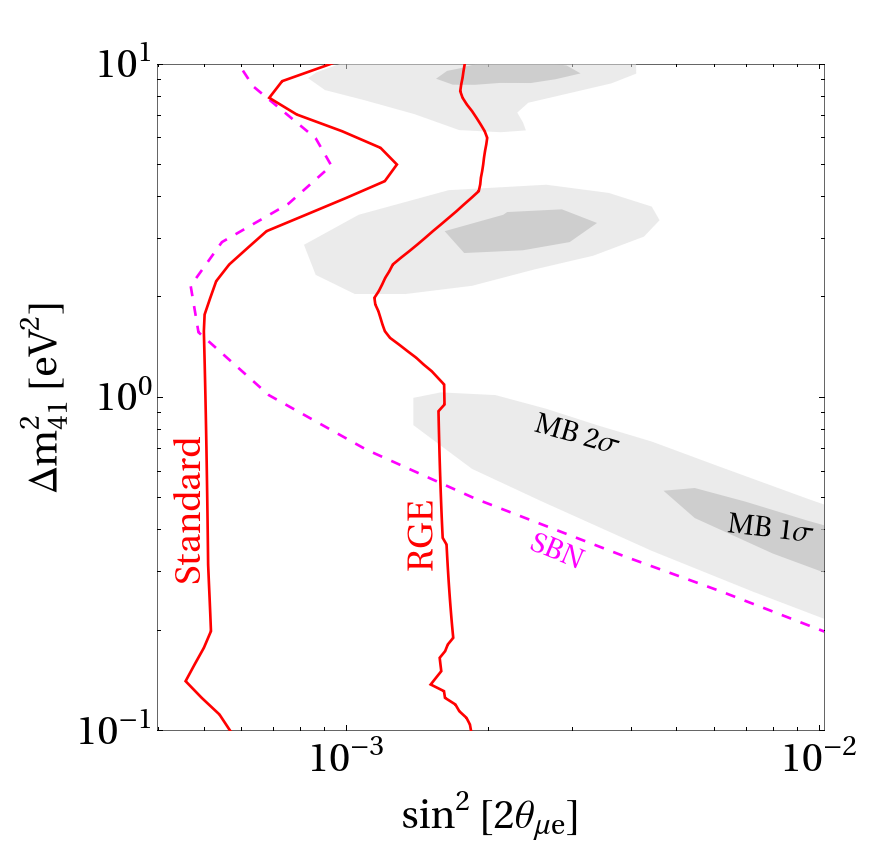} 
	\caption{`Standard:' Combined constraints on $\sin^22\theta_{\mu e}$ as a function of $\Delta m^2_{41}$, assuming the Standard 3+1 scenario. `RGE:' Combined constraints on the effective $\sin^22\theta_{\mu e}$ at MiniBooNE as a function of $\Delta m^2_{41}$ in the model of interest. We include bounds from solar experiments (using the AGSS09 model), MicroBooNE, and $\nu_{\mu}$ disappearance experiments, including MINOS. The region of parameter space preferred by MiniBooNE is depicted in grey. The dashed, purple line is the 3~$\sigma$ CL sensitivity of the SBN Program to $\nu_e$ appearance. See text for details.
}
\label{fig:1}
\end{figure}

\cref{fig:1} depicts in grey the region of parameter space preferred by MiniBooNE at the one sigma and two sigma levels. The contour labelled ``Standard'' indicates the region of parameter space where the constraints from solar data, in the context of AGSS09 model, are saturated at the two sigma level along with the 90\% CL constraint from MINOS. This illustrates the well known mismatch between the appearance and disappearance data -- the ``Standard'' bound has no overlap with the region of parameter space preferred by MiniBooNE.
Furthermore, we note that the ``Standard'' curve only has marginal overlap with the $3\sigma$ CL  $\nu_e$ appearance sensitivity of the Short-Baseline Neutrino Program \cite{Machado:2019oxb,MicroBooNE:2021ktl,MicroBooNE:2021zai} depicted with the dashed, purple line labelled ``SBN'' (from \cite{Arguelles:2021meu}). 

The contour labelled ``RGE'' indicates the region of parameter space where the same constraints are saturated in the model presented here, where $\sin^22\theta_{\mu e}$ stands for the effective $\sin^22\theta_{\mu e}$ defined above. For each value of $M(\mu_0)=\sqrt{\Delta m^2_{41}}$, we performed a dense scan of the $\mu_\mu$ and $g'(\sqrt{Q^2}=5\,\text{MeV})$ parameter space identifying the maximal effective $\sin^22\theta_{\mu e}$ for which all constraints are satisfied. Typically, the scan preferred $g'(\sqrt{Q^2}=5\,\text{MeV})\in[1.35,1.5]$, which guarantees the absence of Landau poles below $10^{6}$ GeV.

Atmospheric neutrino experiments and MINOS yield the most important constraints in the framework with energy-dependent mixing parameters at  $\Delta m_{41}^2\lesssim 1 \, \text{eV}^2$. 
At higher values of $\Delta m_{41}^2$, those limits are somewhat relaxed and MicroBooNE constraints on $\sin^2 2\theta_{\mu e}$ \cite{Arguelles:2021meu} dominate the shape of ``RGE'' curve in \cref{fig:1}. 
Running effects do not really modify the MicroBooNE limits since both MicroBooNE and MiniBooNE use the same beamline and hence the production and detection energy scales are essentially the same.

Relative to the ``Standard'' curve, the ``RGE'' allowed region is significantly shifted to the right and, in this case, there is an overlap between disappearance bounds and the region preferred by MiniBooNE at less than two-sigma. It is clear that when one allows the active--sterile mixing parameters to ``run,'' the sterile-neutrino hypothesis provides a much better fit to the MiniBooNE data combined with constraints from the disappearance of $\nu_e$ and $\nu_{\mu}$ from solar experiments and MINOS, respectively. In the near future, SBN data -- dashed, purple curve -- will be able to decisively explore the region of the new-physics parameter space where there is more agreement between MiniBooNE and the rest of the neutrino data. SBN is well positioned to make a discovery or exclude (or at least severely constrain) the new physics discussed here.

\begin{figure}[t]
	\centering
	\includegraphics[scale=0.375]{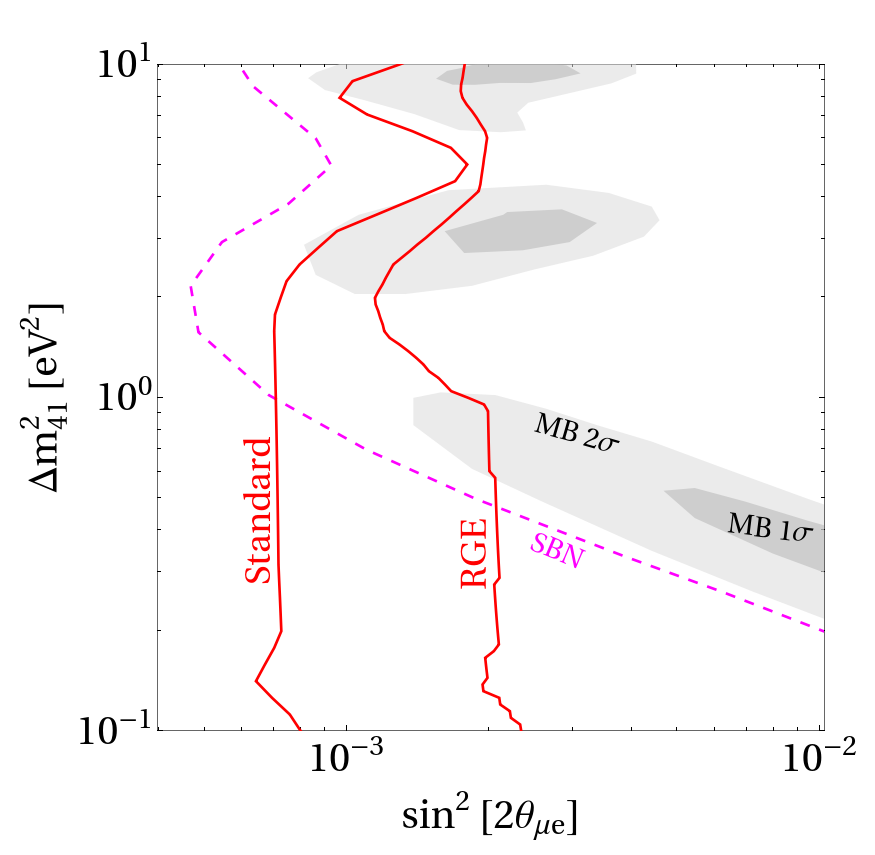} 
	\caption{Same as \cref{fig:1}, except the GS98 solar model is used to extract the bounds from solar data.}
	\label{fig:1a}
\end{figure}

\cref{fig:1a} is similar to \cref{fig:1}, assuming instead the GS98 solar model. This model leads to a less severe solar-neutrino bound on $\theta_{14}$ (at several MeV) and both the ``Standard'' and ``RGE'' curves are shifted to larger (effective) $\sin^22\theta_{\mu e}$ values relative to those in \cref{fig:1}. 

\begin{figure}[t]
	\centering
	\includegraphics[scale=0.375]{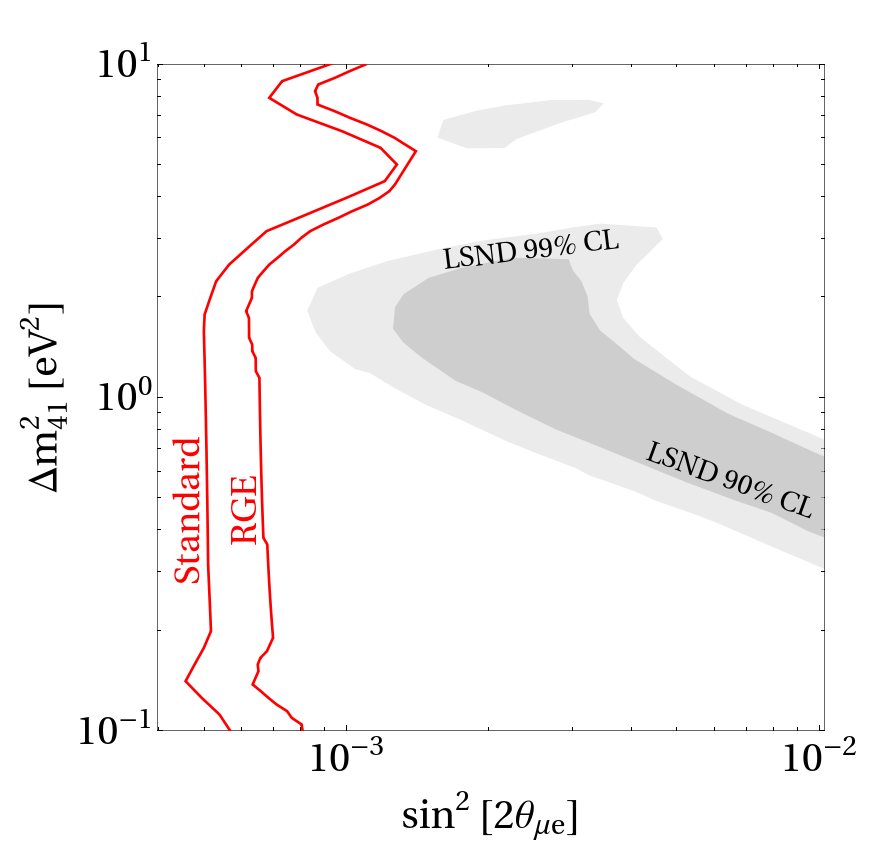} 
	\caption{Same as \cref{fig:1}, for LSND instead of MiniBooNE. The region of parameter space preferred by LSND is depicted in grey. Detection energies are smaller at LSND and the distinction between the two red curves is not as pronounced as in the MiniBooNE case (\cref{fig:1}).}
	\label{fig:1b}
\end{figure}

\cref{fig:1b} is similar to  \cref{fig:1}, except that we concentrate on the LSND anomaly. Relative to  \cref{fig:1}, there are two important changes. One is that the region of parameter space preferred by the LSND data is not identical to that preferred by MiniBooNE. More important is that the detection energy scales associated to LSND are  an order of magnitude smaller than the ones associated to MiniBooNE. Hence, running effects are not as pronounced, and the ``RGE'' curve is similar to the ``Standard'' one. In  a nutshell, even when running effects are taken into account, we do not expect a good combined fit to LSND and short-baseline disappearance experiments.  

\textbf{Summary and Conclusions.} 
The sterile-neutrino interpretation of the MiniBooNE and LSND anomalies is strongly disfavored by other neutrino experiments, especially those that constrain $\nu_e$ and $\nu_{\mu}$ disappearance at short baselines \cite{Dentler:2018sju,Gariazzo:2017fdh,Diaz:2019fwt}. We showed that the situation is significantly improved if the active--sterile mixing parameters depend on the energy scales associated to neutrino production and detection. This can be realized, for example, if there are two sterile neutrinos that couple to a new, light $U(1)$ gauge boson with equal and opposite charges. In this case, RG effects lead to active--sterile mixing parameters that grow as a function of the renormalization scale and strong constraints from solar and reactor data -- neutrino production and detection at 10~MeV -- translate into relatively weaker constraints on the effective mixing parameter that characterizes $\nu_{\mu}\to \nu_e$ oscillation at MiniBooNE and LSND -- neutrino production at 100~MeV, neutrino detection at several hundred MeV. This remains true after properly accounting for higher-energy constraints from short-baseline $\nu_{\mu}$ disappearance. Figs.~\ref{fig:1}, \ref{fig:1a}, \ref{fig:1b} are meant to illustrate that one expects a better fit to all short-baseline data once the running mixing-angle effects are taken into account. They do not reveal whether a global fit to all neutrino data is satisfactory. A thorough combined analysis of all neutrino oscillation data, taking into account the energy-scale dependency of all neutrino mixing parameters, beyond the aspirations of this letter, is required for that. 

\textbf{Acknowledgements.}
This work was supported in part by the US Department of Energy (DOE) grant \#de-sc0010143 and in part by the NSF grant PHY-1630782.
Fermilab is managed by the Fermi Research Alliance, LLC (FRA), acting under Contract No.\ DE-AC02-07CH11359.

\bibliography{refs}

\end{document}